\documentclass[runningheads]{llncs}
\usepackage[T1]{fontenc}
\usepackage{graphicx}
\usepackage{color}
\usepackage{hyperref}
\usepackage{xurl}
\usepackage{paralist}
\usepackage{fontawesome5}
\usepackage{booktabs}
\usepackage{caption}
\usepackage{array}
\usepackage{geometry}
\usepackage[flushleft]{threeparttable}
\begin{document}
\title{Challenges in designing research infrastructure software in multi-stakeholder contexts}
\titlerunning{Research software in multi-stakeholder contexts}
%
\author{Stephan Druskat\inst{1}\orcidID{0000-0003-4925-7248} \and
Sabine Theis\inst{1}\orcidID{0000-0002-3422-3734}}
\authorrunning{S. Druskat \& S. Theis}
%
\institute{
Institute of Software Technology, German Aerospace Center (DLR),\\\{Berlin, Cologne\}, Germany\\
\email{\{stephan.druskat,sabine.theis\}@dlr.de}
}

\maketitle              
\begin{abstract}
This study investigates the challenges in designing research infrastructure software for automated software publication in multi-stakeholder environments, focusing specifically on the HERMES system. Through two quantitative surveys of research software engineers (RSEs) and infrastructure facility staff (IFs), it examines technical, organizational, and social requirements across these stakeholder groups. The study reveals significant differences in how RSEs and IFs prioritize various system features. While RSEs highly value compatibility with existing infrastructure, IFs prioritize user-focused aspects like system usability and documentation. The research identifies two main challenges in designing research infrastructure software: (1) the existence of multiple stakeholder groups with differing requirements, and (2) the internal heterogeneity within each stakeholder group across dimensions such as technical experience. The study also highlights that only half of RSE respondents actively practice software publication, pointing to potential cultural or technical barriers. Additionally, the research reveals discrepancies in how stakeholders view organizational aspects, with IFs consistently rating factors like responsibility structures and quality assurance as more important than RSEs do. These findings contribute to a better understanding of the complexities involved in designing research infrastructure software and emphasize the need for systems that can accommodate diverse user groups while maintaining usability across different technical expertise levels.
\keywords{Research infrastructure software  \and Requirements elicitation \and Multi-stakeholder contexts.}
\end{abstract}
\section{Introduction}

Computing is now regarded as the third pillar of scientific research,
complementing theory and experiment~\cite{skuse_third_2019}.
Consequently, software -- as a central computing component --
has become integral to research.
As such, it fulfills different functions, such as
\begin{inparaenum}[(1)]
    \item research paradigm and method in modeling and simulation,
    \item research tool in computational science and
    \item research output wherever it is created for potential reuse,
        particularly in computer science and related technology research 
        where software prototypes are research results in themselves~\cite{hasselbring_multi-dimensional_2024}.
\end{inparaenum}

On these grounds, modern good scientific practice stipulates that
software shall be treated on par with other research outputs~\cite{jay_software_2021}.
This means that it must be made available for review by peers,
and that it must be cited like any other form of research output~\cite{smith_software_2016}
to provide the research provenance that is necessary to reproduce research results (cf.~\cite{deutsche_forschungsgemeinschaft_guidelines_2022,us_national_science_foundation_proposal_2024}).

These requirements create a necessity for formal software publication~\cite{druskat_towards_2023},
in analogy to the publication of research results and research data.
Software publication activity completes
iterations of the research software cycle, on the one hand,
where software artifacts that have been developed
are being made available in a way that enables
review and reuse by peers,
and the research cycle, on the other hand,
where research artifacts are being made available
for review and reuse by peers,
with software being one type of artifact.

Software publication can be understood as a socio-technical system where humans, technology, and organizational frameworks interact. As such, software systems categorized 
as \textit{research infrastructure software} are not standalone entities; they must address human needs and requirements to support them in achieving their goals and completing tasks efficiently and effectively.  
Without systematic user requirement elicitation and research, technology risks being poorly integrated into these contexts, resulting in inefficiency of the process or organization and frustration or even rejection and non-utilization on an individual level. For this purpose, IEEE Std 29148-2018 requires a systematic elicitation, analysis specification, validation, and management of requirements for a software system, whereby user research serves to identify functional requirements, quality characteristics, and usage conditions. 
It is initially essential to look at the type of technology considered for eliciting requirements. 

In general, research infrastructure software to support software publication includes different software systems:

\begin{itemize}
    \item \textit{Collaborative software development platforms (CSDP)} for working together on a shared codebase, managing changes, and organizing collaboration. These platforms combine version control systems with bug or issue trackers, knowledge bases (wikis, static websites), project management tools (e.g., kanban boards), messaging capabilities, etc. Examples include GitLab, GitHub, Forgejo, etc.
    \item \textit{Continuous integration and deployment (CI/CD) systems}
    for automating software development tasks such as testing, building, releasing, etc.
    These use virtual software environments that users can configure to run everyday tasks
    based on specific triggers, e.g., new code versions being pushed to a CSDP.
    \item \textit{Software systems or packages providing automation tasks (ATS)}.
    These are systems that are run within the virtual environments or orchestrated by a CI/CD system
    and execute automation tasks.
    Examples include language-specific libraries for software testing, static code analysis platforms, and automated software publication workflow systems.
    \item \textit{Publication repository platforms (PRP)} for depositing records of software metadata and artifacts.
    PRPs archive any deposited artifacts, record and display metadata about records,
    and provide search or browsing functionality.
    Examples include Invenio/InvenioRDM, Dataverse, DSpace, EPrints, etc.
\end{itemize}

\noindent Any of these software systems have multiple stakeholder groups in the context of research.
Minimally, we can distinguish between 
\textit{users} of the system, 
i.e., developers of software within research who want to publish their software;
\textit{developers} of research infrastructure software as a specific category of software that is being developed in research (see~\cite{hasselbring_multi-dimensional_2024} and section~\ref{subsec:background:categories});
\textit{operators} of research infrastructure software, 
i.e., infrastructure facility staff at research organizations who provide research infrastructure software and systems to their users, including researchers and developers.

All stakeholder groups place different requirements on research infrastructure software, which complicates the design of research infrastructure software. Where research infrastructure software is designed and developed in an academic setting,
this is exacerbated by a lack of focus on and resources for software and systems development (see section~\ref{subsec:background:rseng}).
This article contributes to a better understanding of the challenges of designing and developing research infrastructure software.
We present a case study highlighting 
the differing requirements by the stakeholder groups
encountered for a specific research infrastructure software ATS, \textit{HERMES}.
HERMES~\cite{druskat_software_2022}
provides a configurable and extensible workflow for automating software publication with rich metadata in CI/CD systems.
It allows its users to publish software in a way that satisfies the FAIR Principles for Research Software (FAIR4RS~\cite{chue_hong_fair_2021,barker_introducing_2022}).
The specific challenges faced for the design and development of this particular system relate to
multiple stakeholder groups and their intrinsic heterogeneity.
We conducted online surveys with research software engineers (RSEs, see section~\ref{subsec:background:rseng})
and staff at research infrastructure facilities 
to gather insights into each group’s attitudes
and requirements towards HERMES.
Our results show that both the different roles of stakeholders, as well as each group’s heterogeneity, lead to
specific challenges for the design of HERMES.

\section{Background}

The following sections briefly introduce related central aspects
to establish an understanding of the context for the study presented here.

\subsection{Research Software Engineering}
\label{subsec:background:rseng}

\begin{figure}
    \centering
    \includegraphics[width=.9\textwidth]{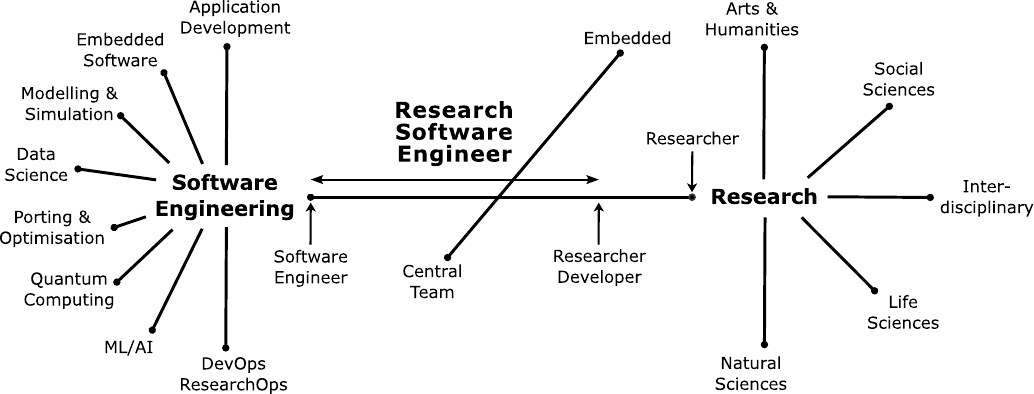}
    \caption{Dimensions of diversity in the research software engineer (RSE) role (adapted from~\cite{druskat_software_2024})}
    \label{fig:rse-role}
\end{figure}

\textit{Research software} is defined as including ``source code files, algorithms, scripts, computational workflows and executables that were created during the research process or for a research purpose''~\cite[p. 16]{gruenpeter_defining_2021}.
\textit{Research software engineering} is the practice of applying software engineering methods
to the creation and maintenance of research software.
As such, it is practiced in academic research by \textit{research software engineers} (\textit{RSEs}).
The RSE role has emerged over the last decade to include people who focus on software development and maintenance in a research setting.
The role is inclusively defined, and represents a spectrum of experiences and skill sets with regard to software engineering~\cite{dworatzyk_decoding_2024} (see~\autoref{fig:rse-role}).
RSEs are also heterogeneous in their activities in different areas of software engineering and the application of software engineering methods.
Finally, RSEs often have an academic background in a specific academic discipline themselves, but are usually not trained software engineers~\cite{hettrick2022RSESurvey}.

The work context in which research software engineering is most often practiced is that of academic research.
In this context, ubiquitous fixed-term work contracts and the process of academic qualification
in combination with the lack of formal education in software engineering
lead to specific constraints on research software engineering practice in terms of resources (e.g., team sizes)
and the applicability of typical software engineering practices as detailed in the Guide to the Software Engineering Body of Knowledge (SWEBOK, \cite{washizaki_guide_2024}).

Research software engineering is also concerned with additional requirements specific to the academic work area.
Modern guidelines for safeguarding good research practice (e.g.~\cite{deutsche_forschungsgemeinschaft_guidelines_2022}), as well as funders' principles for handling research software (e.g.~\cite{deutsche_forschungsgemeinschaft_handling_2024,us_national_science_foundation_proposal_2024}),
mandate for publicly funded software to be made available following the FAIR Principles for Research Software (FAIR4RS, \cite{chue_hong_fair_2021}, see~\autoref{subsec:background:fair4rs}).
Research software must also be made citable, and cited~\cite{smith_software_2016}.
Therefore, research software engineering also includes additional activities concerned with adhering to these principles,
beyond those documented in SWEBOK:
metadata creation, provision, and maintenance; 
implementation of software publication (see~\autoref{subsec:background:software-publication});
implementation of software interoperability, e.g., through the use of standard formats and domain-specific data schemas, or the provision of APIs;
safeguarding reusability through the conservation of runtime environments, modularization, etc.;
software documentation for stakeholders, including a description of research domain-specific properties of the software.

\subsection{Research Infrastructure Software (RIS)}
\label{subsec:background:categories}

Research software is not a homogeneous class of software.
Hasselbring et al.~\cite{hasselbring_multi-dimensional_2024} define categories of research software across multiple dimensions:
the software's role in research, the software's technology readiness level, the developers of the software, the dissemination of the software.
For software's role in research, three subcategories are defined: 
modeling, simulation, and data analytics software; 
technology research software; 
research infrastructure software.
Of those, this paper is concerned with research infrastructure software.

While modeling, simulation, and data analytics software (Cat. 1) is created and applied to answer domain research questions directly,
and technology research software (Cat. 2) is developed to provide proofs for technological concepts,
research infrastructure software (Cat. 3, \textit{RIS}) supports research.
Examples include control and monitoring software,
data collection and data generation software,
pipelines and related tools,
software libraries (e.g., for high-performance computing),
management software for research artifacts, etc.
Notably, software in Cat. 1 and Cat. 2 can evolve to become RIS~\cite{hasselbring_multi-dimensional_2024}.

RIS usually exhibits a technology-readiness level $>= 5$~\cite{european_commission_directorate_general_for_research_and_innovation_technology_2017}.
Therefore, the development of RIS requires quality assurance methods that do not necessarily apply to other research software, e.g., 
requirements engineering or safety analysis.
RIS has a set of stakeholders that encompasses its researcher end users (same as for Cat. 1 and Cat. 2), but also 
infrastructure operators, infrastructure support and training staff, 
and potentially others, such as organizational leadership, quality management, etc.
This is inherent to RIS' role as infrastructure that is required to support the mission of the research organization,
adhere to legal standards,
support good research practice,
be cost-effective to operate,
and be interoperable with other infrastructure.

\subsection{FAIR Principles for Research Software}
\label{subsec:background:fair4rs}

The FAIR Principles for Research Software (FAIR4RS)~\cite{chue_hong_fair_2021} apply
the FAIR Principles for research data~\cite{wilkinson_fair_2016} (Findability, Accessibility, Interoperability, Reusability)
to research software to improve its sharing and reuse.
Software is \textit{findable}, when it and its associated metadata are easy for humans and machines to find.
Software is \textit{accessible}, when software and its metadata are retrievable via standardized protocols.
Software is \textit{interoperable}, when it interoperates with other software by exchanging data/metadata and/or 
providing APIs, described through standards.
And software is usable when it can be executed, and \textit{reusable} when it can be understood, modified, built upon and/or incorporated into other software.

\subsection{Software Publication}
\label{subsec:background:software-publication}

Software publication is the ``process of depositing software (versions) and metadata in a publication repository, where the deposit is uniquely identified with a machine-resolvable persistent identifier''~\cite[p.~169]{druskat_towards_2023}.
As such, software publication partially fulfills FAIR4RS principles.
Through software publication, the software and its metadata are easy to find for both humans and machines.
Publication repositories let users retrieve software (metadata) records through standardized protocols.
Published metadata makes software more understandable, which contributes to software reusability.

A more complete fulfillment of FAIR4RS principles additionally needs research software engineering (see~\autoref{subsec:background:rseng}),
through which software can be made interoperable, executable, and more reusable.

\subsection{HERMES Workflows for Automated Research Software Publication}

\begin{figure}
    \centering
    \includegraphics[width=.9\textwidth]{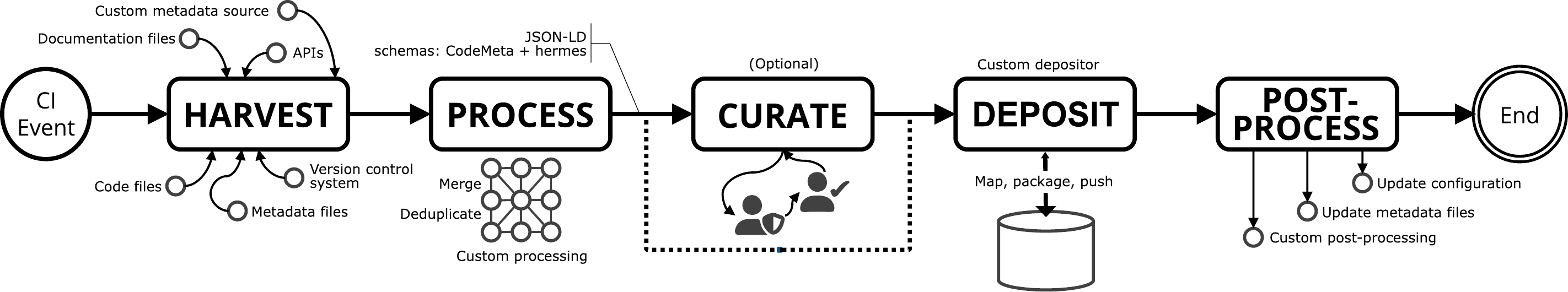}
    \caption{HERMES workflow phases}
    \label{fig:hermes}
\end{figure}

Software publication with metadata that fulfills FAIR4RS principles as outlined above requires
the creation, compilation, and transmission of metadata pertaining to the published software (version)
to the publication repository.
This is often a manual task, e.g., where web forms must be completed for each new publication.
RIS can help optimize the software publication process through automation.
HERMES~\cite{druskat_software_2022} is a workflow for automated publication of software with rich metadata.
It is prototyped in the \texttt{hermes} Python package~\cite{meinel_hermes_2024} and uses CI/CD systems
to automatically harvest and process existing software metadata, lets them be curated and signed off on, 
and deposits them in publication repositories together with software artifacts~\cite{kernchen_extending_2024}.
Optionally, the metadata can be post-processed, e.g., fed back into the software source code repository.
\autoref{fig:hermes} outlines the HERMES workflow phases.

Currently, HERMES has successfully been validated in lab conditions~\cite{kernchen_extending_2024}.
Developing the HERMES prototype into research infrastructure software requires its validation and demonstration in a relevant environment (see~\cite{european_commission_directorate_general_for_research_and_innovation_technology_2017}),
which in turn requires careful system design.
Relevant environments for HERMES are research organizations.
A successful demonstration of HERMES in research organizations includes its
successful application by users, i.e., RSEs and researchers at the organization,
as well as HERMES' operation and support through the organization's
infrastructure provider and support staff.
For the present study, we define this environment as a multi-stakeholder context.

\section{Problem and Research Questions}

Designing research infrastructure software for software publication is inherently challenging in multi-stakeholder contexts. Stakeholders such as RSEs and infrastructure facility staff (IF) place diverse and sometimes conflicting requirements on these systems. This complexity is further compounded by the work context, e.g., in academic institutions or libraries. Here, limited resources and a lack of focus on software development often constrain the design and implementation of practical solutions. So far, it remains unclear to which extent existing tools, such as collaborative software development platforms (e.g., GitHub), CI/CD systems, and publication repositories, address the multifaceted needs of RSEs. The development of automated workflows, such as those provided by HERMES, may be hindered by barriers that emerge from the heterogeneity of stakeholder requirements, ranging from technical functionalities to organizational and social factors. Understanding these contexts and challenges is critical for designing research infrastructure software that aligns with the FAIR Principles for Research Software (FAIR4RS) and supports the effective dissemination of research outputs.
The present study therefore aims to address the following research questions:
\begin{enumerate}
    \item[(1)] What are the technical, organizational, and social challenges faced in designing research infrastructure software for automated software publication?
    \item[(2)] How do stakeholders, such as RSEs and infrastructure facility staff, differ in their requirements and expectations for research infrastructure software?
\end{enumerate}
Study results will be able to provide a basis for understanding and addressing the multidimensional challenges associated with automatic publication of research software. 
They will provide a first glance into the design of systems that facilitate reproducibility, collaboration, and efficiency in research software publication. 

\section{Method}

To answer our research questions, we conducted two quantitative user surveys, one to elicit the requirements of infrastructure facility staff (IF) and the other to elicit the requirements of research software engineers (RSE).
The surveys were implemented as online questionnaires using an instance of LimeSurvey~\cite{limesurvey_gmbh_limesurvey_nodate} hosted at the German Aerospace Center.

\subsection{Participants}
    \subsubsection{RSE.} In total, N=83 participants filled in the questionnaire. The sample consisted of primarily male (80.7\% ) persons to the larger part between 35 and 44 (51.8\%) years old. Respondents had a professional background predominantly in natural sciences (39.8\%), computer science (25.3\%), and engineering (15.7\%). From 83 participants, 1.2\% (N=1) stated that they had no experience in research software engineering. 25.3\% (N=21) of the participants reported that they have professional experience of 1 to 3 years, while 20.5\% (N=17) had 4 to 6 years, and another 20.5\% (N=17) had 7 to 10 years of experience. The largest group, 32.5\% (N=27) of the participants, had more than 10 years of professional experience.
    \subsubsection{IF.} N=39 participants answered the survey to investigate the requirements and characteristics of infrastructure facility staff (IF). Among them, 18 participants (46.2\%) identified as male, and 10 participants (25.6\%) identified as female. Additionally, 3 participants (7.7\%) identified with other genders, while 8 (20.5\%) chose not to disclose their gender. Three participants classified themselves (7.7\%) under 25 years old. The largest age group was 35 to 44 years, with 14 participants (35.9\%). 9 participants (23.1\%) were in the 25 to 34, 6 participants (15.4\%) in the 45 and 54 years ranges, and 7 participants (17.9\%)  in the 55 to 64 age group. Participants indicated that they primarily work in university libraries (43,6\%), with a majority (61.8\%) concerned with the operation, and around a third with development (35.3\%) of infrastructure components such as open access or source code repositories. Other tasks included advising users on data management (61.8\%) and managing research data (52.9\%) or repositories (58.8\%).
    \subsection{Survey Design}
    We describe both surveys in each subsection of this methodology.
    \subsubsection{RSE.} The questionnaire for RSEs' requirements contained 14 questions of four thematic sections,  addressing aspects of the working environment, functional requirements, organizational and social aspects, individual aspects, and capacity building.
    The first section queried participants' work-related background and experience, including the duration of their work as RSEs and their role as RSEs or software developers. Furthermore, a question on experience in terms of working years was included, with response options ranging from ``No experience'' to''>10 years.'' This section established the relevance of participants' expertise to the study's objectives.
    The second section contained questions on RSEs' perception of the technical functionalities of software publication workflow tools. Answer options included a multiple-choice list of various technical features, such as configurability, interoperability with other systems, compliance with metadata standards, and availability of tutorials and documentation. Participants rated these technical features on a four-point Likert scale from ``very important'' to ``not important.'' Additionally, a multiple-choice question prompted participants to select technical functions essential for HERMES integration, such as automatic metadata updates, support for unstructured metadata sources, and compatibility with continuous integration systems.
    The third section addressed organizational and social aspects related to the use of HERMES workflows. Questions focused on responsibility structures for software maintenance, community management, and quality assurance procedures. A Likert scale captured participants' priorities in these areas. Furthermore, participants then evaluated the clarity of collaboration between RSEs and researchers on a dichotomous scale (``Clear and defined'' vs. ``Unclear and vague''). They identified measures to enhance this collaboration, including workshops, improved communication platforms, and clear role definitions.
    The fourth section contained questions about individual aspects and user needs for capacity building. Answer options included a multiple-choice list, including introductory courses, advanced workshops, online tutorials, and discussion forums. This list also included the option to add ``other'' training options in an open text field.
    The last section contained demographic questions, such as gender, age group, and field of research, and an open-ended question to share additional comments or suggestions for improving HERMES. The questionnaire took approximately 10 minutes to complete. All answer options and questions resulted from discussion and iterative feedback with RSEs and human factors experts from the DLR Institute of Software Technology.
    \subsubsection{IF.} The IF survey was similar but did not completely overlap with the RSE one. It contained 25 questions divided into work context, requirements, organizational and social aspects, individual characteristics and capacity-building needs, and demographic questions about the individual participant.
    In the first section querying information on the work context, participants indicated the type of infrastructure facility they work in, such as, e.g., libraries, research data management, or IT services. Then, they had to specify their roles, including, e.g., repository managers, legal advisors, or technical support staff. While there was only one answer to choose from a list of options regarding the type of institution, they could select multiple roles. In both cases, an open text field allowed adding an answer the list did not contain.
    Additionally, participants provided details about the infrastructure components their organization manages or supports, including open access publication repositories (e.g., DSpace), source code repository platforms (e.g., GitHub), and continuous integration tools (e.g., Jenkins) via open text fields and by choosing multiple options from a list. Furthermore, participants had to answer by dichotomous answer option (yes, no) whether institutional open access repositories have a dedicated input type for software or provide opportunities for software publication, and if they contain published software. Further items in this section included whether guidelines are available for managing research software and the types of guidelines provided (e.g., development, licensing, publication, or management). Further inquiries addressed whether institutions offer advice, training (either directly or through documentation), or support via RSEs (centralized, external, or community-based). Data collection used nominal or categorical scales, with conditional logic ensuring relevance by displaying items only when meeting prior criteria.
    Compared to the RSE questionnaire, the section on functional requirements of the users of the IF questionnaire was relatively short, only containing one question, where respondents evaluated the significance of various aspects using a four-point Likert scale: significant, meaningful, less important, or unimportant. These aspects included the system's usability for researchers, its adaptability within infrastructure facilities, the availability of open-source licensing, cost-free access, comprehensive documentation for users, administrators, and developers, implementation in a specific programming language, compatibility with existing infrastructure components, ease of integration, adherence to metadata standards, availability of templates for training materials and self-learning resources for users and administrators, and the inclusion of administrator training as part of the infrastructure facility.
    Subsequently, the third section evaluated the importance of specific organizational aspects related to using a software system for automated software publication. Participants rated each element on a four-point Likert scale, ranging from very important to unimportant. The organizational aspects assessed included responsibility structures for maintaining the software system, the availability of rules and guidelines for its use, the implementation of community management and building efforts, and the adoption of quality assurance procedures. Respondents subsequently evaluated the role comprehension of researchers and RSEs on a dichotomous scale (``clear and defined'' versus ``unclear and vague''). Using multiple-choice options, they could also suggest ways to improve collaboration, such as workshops, communication platforms, or more explicit role definitions.
    The fourth section addressed individual capacity-building and training needs. Participants here initially indicated the training and educational measures they considered necessary to ensure the proper use of an automated software publication system by its users. Respondents could select multiple options, including introductory courses on platform usage, advanced workshops on specific features, online tutorials and documentation, forums and discussion groups for experience exchange, intensive courses such as summer or winter schools, and an open-ended ``other'' option to specify additional suggestions. They also proposed additional training needs in open-text fields. The following question explored the training and educational measures considered necessary to facilitate the proper setup, configuration, administration, and operation of a system for automated software publication within infrastructure facilities. Respondents could select multiple options, including introductory courses on platform usage, advanced workshops on specific features, online tutorials and documentation, forums and discussion groups for experience exchange, intensive courses such as summer or winter schools, and an open-ended ``other'' option to specify additional measures.
    The final section included questions on demographic details such as gender, age, and field of research. Participants completed the survey in approximately 15 minutes. All questions were developed in collaboration with an expert in software publication and research software engineers to ensure their relevance to the target audience. 
 \subsection{Data Analysis}
    The statistical analysis of both closed questions in both questionnaires was descriptive, including frequencies and percentages for categorical answers and standard deviations and mean values for Likert scale responses. We coded open-ended responses thematically to identify frequently occurring options. All statistical analyses were conducted using SPSS (version 26). Figures 3-6 were produced using Pandas (Version v2.2.3~\cite{the_pandas_development_team_pandas_2024}) and Matplotlib~\cite{hunter_matplotlib_2007} (Version v3.10.0~\cite{the_matplotlib_development_team_matplotlib_2024}).

\section{Results}

\subsection{Work Context}
\subsubsection{RSE.} Half of the respondents (50.6\%) publish every version of their research software in, e.g., an open-access repository. At the same time, a smaller group of 20.5\% of the sample mentioned not publishing their research software. This emphasizes promoting FAIR4RS principles and making software publication processes and tools more user-friendly. Other participants reported one-time publication methods, such as through a scientific article (9.6\%), a software paper focusing on source code and documentation (4.8\%), or a single upload to an open-access repository (13.3\%). Of the participants who indicated that they publish their research software in an open-access repository, the majority (53\%) did so in a general-purpose one such as Zenodo. A smaller proportion (8.4\%) used a repository provided by their institution, while 2.4\% used a discipline-specific repository. These results show that publicly available general-purpose repositories are used predominantly and should, therefore, be specifically supported by automated software publication. Participants who used open-access repositories (N=83) also provided information regarding how they publish their research software. 4.8\% of them manually prepare their open-access software publications and intend to maintain this approach.
In contrast, 32.5\% stated that while they currently prepare their publications manually, they would prefer to automate the process. 26.5\% of respondents have adopted existing automation tools for publishing their software in open-access repositories. From N=22 participants who already use automated software publication tools, 21 indicate that they use the Zenodo-GitHub integration, and only one participant used their own Python scripts in CI/CD or HERMES. 
\begin{table}[ht]
\centering
\caption{Usage of technical infrastructure components}
\begin{tabular}{@{}lccc@{}}
\toprule
Component & N & \% Responses & \% Cases \\ 
\midrule
Open Access Publication Repositories      & 35  & 46.7\% & 97.2\% \\
Source Code Repository Platforms          & 25  & 33.3\% & 69.4\% \\
CI/CD Systems                             & 14  & 18.7\% & 38.9\% \\
None of the Above                         & 1   & 1.3\%  & 2.8\%  \\ 
\midrule
Total                            & 75  & 100\%  & 208.3\% \\ \bottomrule
\end{tabular}
\label{tab:infrastructure_components}
\end{table}

\subsubsection{IF.} The institutions in which the infrastructure employees, as a user group for automated software publication, work, have already been mentioned in the sample description. The analysis revealed that the most frequently reported roles among infrastructure facility staff were the operation of infrastructure components (61.8\% of cases), user consulting (61.8\% of cases), and repository management (58.8\% of cases). Research data management was also a significant area of engagement, reported by 52.9\% of cases. Development of infrastructure components was identified in 35.3\% of cases. In contrast, leadership roles, such as management or coordination of infrastructure teams (23.5\% of cases) and leadership of infrastructure facilities (8.8\% of cases), were less common. Specialized roles, including Open Science Officers (5.9\% of cases) and Data Stewards/Software Stewards (11.8\% of cases), were among the least reported. These findings indicate that operational and consultative roles dominate the work of infrastructure facility staff, with comparatively fewer individuals involved in leadership or specialized functions. Regarding the results of the technical infrastructure components among institutions, as well as specific practices regarding software publication and research software management, it needs to be considered that respondents were allowed to select multiple answers.
The majority of institutions operate open-access publication repositories (46.7\% of responses, 97.2\% of cases) and source code repository platforms (33.3\% of responses, 69.4\% of cases). CI/CD systems are less commonly used, with 18.7\% of responses and 38.9\% of cases. Notably, only 2.8\% of cases indicated no use of the listed components.
Among open access repositories, DSpace is the most widely used platform (25.6\% of cases). Other platforms, such as OPUS (5.1\%) and Dataverse (2.6\%), are also employed, but to a lesser extent. Regarding the explicit publication of software, 43.6\% of institutions offer the capability, while 46.2\% have already published software in their repositories.
Regarding guidelines for research software management, 28.2\% of institutions have policies, while 46.2\% reported having no formal guidance. A notable proportion (17.9\%) were uncertain about the existence of such guidelines.
\begin{figure}
    \centering
    \includegraphics[width=.8\textwidth]{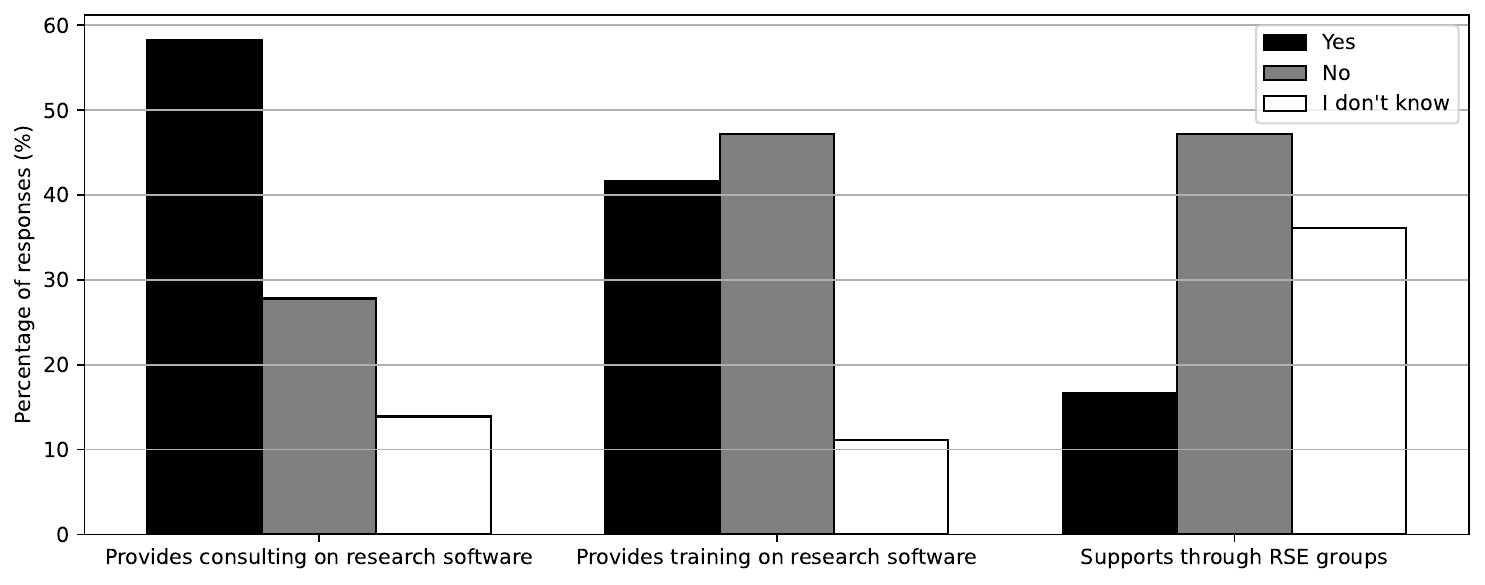}
    \caption{Institutional support for software publication for IFs}
    \label{fig:institutionalsupportIF}
\end{figure}

The survey data reveal varying levels of support provided by infrastructure facilities to users in managing research software (see~\autoref{fig:institutionalsupportIF}). Most institutions (53.8\%) actively provide consulting on research software usage, with a smaller proportion (25.6\%) reporting no consulting services, and 12.8\% unsure about their practices.
In terms of training, only 20.5\% of facilities directly provide training programs or workshops, while 17.9\% support users indirectly through self-learning materials or documentation. However, 43.6\% of institutions offer no training services, and 10.3\% of respondents were unsure.
Support through RSEs remains limited. Only 7.7\% of institutions have a central or self-organized RSE group, while 43.6\% lack such support structures altogether. A significant portion (33.3\%) of respondents were unsure about the existence of RSE-related services within their institutions. Findings show that consulting services are relatively common, but structured training and dedicated RSE groups are still underrepresented in infrastructure facilities, highlighting potential areas for improvement.

\subsection{Requirements}
\subsubsection{RSE.}
As~\autoref{tab:table2} shows, integrating tools for automated software publication with infrastructure components, such as source code repositories, continuous integration tools, and publication repositories, had the highest priority for respondents. 83.1\% of a total of 83 who answered this question considered this function as  ``very important,'' while 14.5\% rated it as ``important,'' resulting in a combined importance score of 97.6\%. Ease of immediate use was another critical aspect, with 88\% of participants rating an applicability ``out-of-the-box'' as ``very important'' and  ``important.'' 80.8\% of the respondents considered the interoperability of automated software publication tools like HERMES as ``very important'' or ``important.'' Adherence to metadata standards was considered a high-priority feature, with a combined importance score of 81.\%. Of the participants, 45.8\% rated it as ``very important,'' and 36.1\% rated it as ``important.'' Clear instructional guides were valued by respondents, with 36.1\% assigning ``very important'' and 45.8\% assigning ``important'' ratings, resulting in a total score of 81.9\%. The presence of ready-to-use templates was rated as ``very important'' by 21.7\% and ``important'' by 48.2\%, achieving a combined score of 69.9\%.
\begin{table}[ht]
    \resizebox{\textwidth}{!}{%
\begin{threeparttable}
    
    \centering
    \caption{RSE's importance ratings of features of automated research software publication systems}
    \label{tab:table2}
    \begin{tabular}{lcccc}
        \toprule
        Aspect & Very important (\%) & Important (\%) & Less important (\%) & Unimportant (\%) \\
        \midrule
        Compatibility with infrastructure systems & 83.1 & 14.5 & 1.2  & 1.2 \\
        Out-of-the-box usability                 & 47.0 & 41.0 & 9.6  & 1.2 \\
        Compliance with metadata standards       & 45.8 & 36.1 & 15.7 & 1.2 \\
        Interoperability with other systems      & 42.2 & 38.6 & 18.1 & 1.2 \\
        Availability of how-to guides            & 36.1 & 45.8 & 15.7 & 1.2 \\
        Availability of reference documentation  & 30.1 & 41.0 & 24.1 & 3.6 \\
        Availability of tutorials                & 28.9 & 33.7 & 32.5 & 3.6 \\
        Technical documentation with explanations & 26.5 & 33.7 & 32.5 & 6.0 \\
        Availability of templates                & 21.7 & 48.2 & 25.3 & 3.6 \\
        Software extensibility                   & 20.5 & 41.0 & 31.3 & 6.0 \\
        API usability as a library               & 16.9 & 37.3 & 31.3 & 13.3 \\
        High configurability                     & 15.7 & 54.2 & 27.7 & 1.2 \\
        Local CLI usability                      & 12.0 & 37.3 & 38.6 & 10.8 \\
        \bottomrule
    \end{tabular}
    \begin{tablenotes}
      \small
      \item \textit{Note.} The table summarizes the distribution of responses from RSEs (\%) for the importance of each technical aspect in an automated software publication workflow. The data are sorted by percentage of ``Very important'' answers in descending order.
    \end{tablenotes}
\end{threeparttable}}
\end{table}

\begin{figure}
    \centering
    \includegraphics[width=\textwidth]{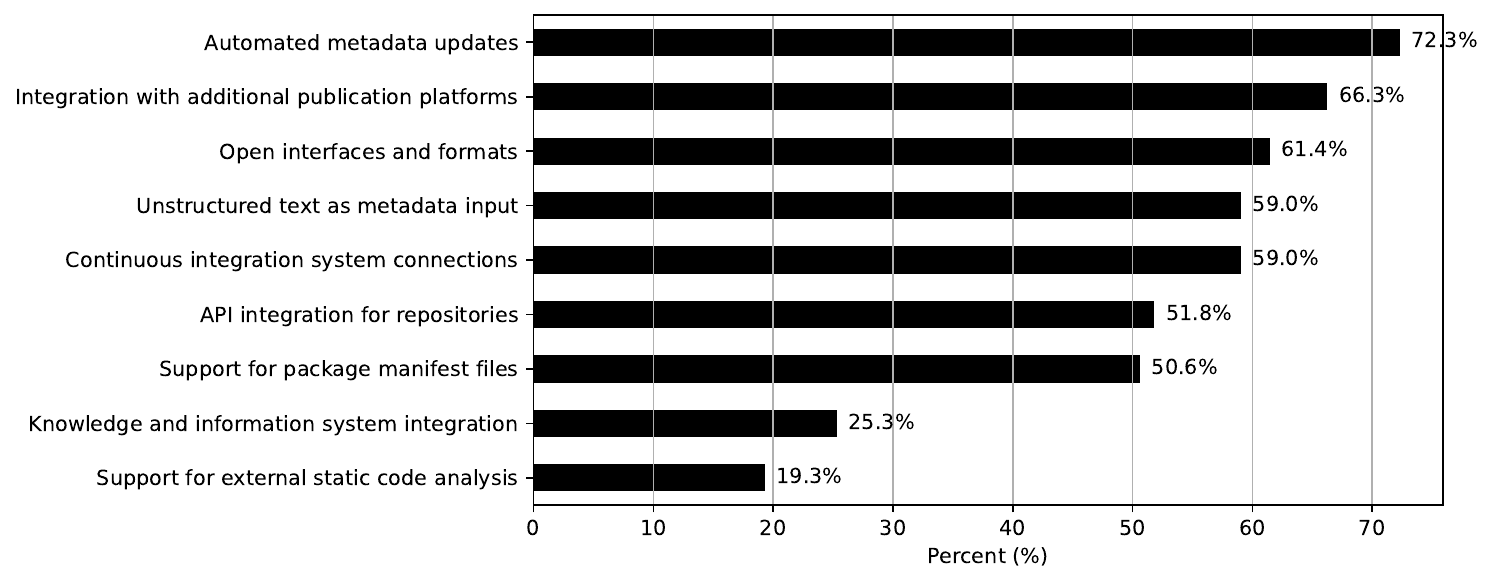}
    \caption{RSE's preferences for technical features in automated software publication systems}
    \label{fig:preferences}
\end{figure}

The results of the survey show which aspects RSEs consider relevant in a system for automated software publication. The highest approval was given to the function “automatic update of metadata in the source code repository,” which was considered necessary by 72.3\% of the respondents. Additionally, 66.3\% of participants rated connectivity to publication platforms as required. Their answers demonstrate the significance of the distribution and visibility of research software. 61.4\% rated ``support for open interfaces and formats'' as necessary. Features such as support for unstructured text as a metadata source (59.0\%) and integration with additional continuous integration systems (59.0\%) also received a high level of agreement. They underscore the need for flexible and integrative solutions that consider standard working methods and infrastructures in software development.
In contrast, support for external platforms for static code analysis (19.3\%) and integration with knowledge management and information systems (25.3\%) received comparatively little approval. This could indicate that these functions are either considered less critical or have been less of a focus in day-to-day practice.
Regarding the technical functions, results demonstrate the relative importance. Automatic metadata updates, integration with additional publication platforms, and open interfaces emerged as the top three features for RSEs. At the same time, features such as support for external static code analysis and integration with knowledge systems were the least prioritized (see~\autoref{fig:preferences}).
In addition to the predefined answers, the participants suggested that the system should create packages (e.g., for platforms like PyPI or conda-forge), enabling both themselves and their colleagues to benefit from standardized installation procedures and versioning. They would like to have a feature to extract metadata from markdown files, such as README.md or CONTRIBUTORS.md, and they proposed the ability of automated software publication systems to reserve DOI (Digital Object Identifiers) and automatically update relevant metadata in the project to ensure proper citation and tracking. The automated software publication system should also avoid complicated steps or administrative access to repositories, making it more accessible and user-friendly.

\subsubsection{IF.} Infrastructure facility staff overwhelmingly prioritized the aspect of high usability for automatic research software publication, with 84.6\% rating it as ``very important''. With 59.0\%, open-source licensing is another critical factor,  identifying it as very important or with 30.8\% as ``important.'' Also highly prioritized is availability for both users (64.1\%) and administrators (59\%), rating it as ``very important''. The results highlight a clear prioritization of usability, open-source licensing, documentation, and compatibility in the design and implementation of automated software publication systems. These factors should be the focal points in future system development to meet user needs effectively. Cost sensitivity emerged as a significant factor, with 33.3\% of participants considering a cost-free system as very important and an additional 35.9\% rating it as important. This underscores the necessity of minimizing financial barriers to adoption. Compatibility and integration also stand out as critical considerations, as 48.7\% of respondents rated compatibility with existing infrastructure and ease of integration as very important, highlighting the importance of ensuring seamless operational functionality. Conversely, the implementation of a specific programming language appears less significant, with 51.3\% of respondents rating it as ``less important'' and 28.2\% considering it ``unimportant''. Lastly, compliance with metadata standards is a prominent concern, with 53.8\% of participants rating it as very important, reflecting a clear emphasis on standardization and interoperability. Technical functions were not requested in the IF questionnaire.

\begin{table}[ht]
\resizebox{\textwidth}{!}{
\begin{threeparttable}
    \centering
    \caption{IF's importance ratings of features of automated research software publication systems}
    \label{tab:table3}
    \small 
    \begin{tabular}{lcccc}
        \toprule
        \textbf{Feature} & \textbf{Very important (\%)} & \textbf{Important (\%)} & \textbf{Less important (\%)} & \textbf{Unimportant (\%)} \\
        \midrule
        High usability for researchers            & 84.6 & 2.6  & 2.6  & 2.6 \\
        Documentation for users                   & 64.1 & 28.2 & 0.0  & 0.0 \\
        Documentation for administrators          & 59.0 & 33.3 & 0.0  & 0.0 \\
        Open source licensing                     & 59.0 & 30.8 & 0.0  & 2.6 \\
        Compliance with metadata standards        & 53.8 & 33.3 & 2.6  & 2.6 \\
        Easy integration into infrastructure      & 48.7 & 35.9 & 5.1  & 2.6 \\
        Compatibility with existing infrastructure & 48.7 & 33.3 & 7.7  & 2.6 \\
        High adaptability to infrastructure       & 23.1 & 51.3 & 17.9 & 0.0 \\
        Documentation for further development     & 30.8 & 41.0 & 17.9 & 2.6 \\
        Cost-free system                          & 33.3 & 35.9 & 23.1 & 0.0 \\
        Templates for user training               & 10.3 & 46.2 & 33.3 & 2.6 \\
        Templates for self-learning (users)       & 7.7  & 48.7 & 33.3 & 2.6 \\
        Templates for self-learning (administrators) & 7.7 & 48.7 & 33.3 & 2.6 \\
        Training for administrators               & 7.7  & 43.6 & 35.9 & 5.1 \\
        Implementation in a specific language     & 0.0  & 12.8 & 51.3 & 28.2 \\
        \bottomrule
    \end{tabular}
    \begin{tablenotes}
      \small
      \item \textit{Note.} The table summarizes the distribution of responses from IF (\%) for the importance of each feature in an automated software publication system. The data are sorted by percentage of ``Very important'' answers in descending order.
    \end{tablenotes}
    \vspace{0.5cm}
    
\end{threeparttable}}
\end{table}

\subsection{Organizational and Social Factors}
\subsubsection{RSE.}
\begin{figure}[h!]
    \centering
    \includegraphics[width=.8\textwidth]{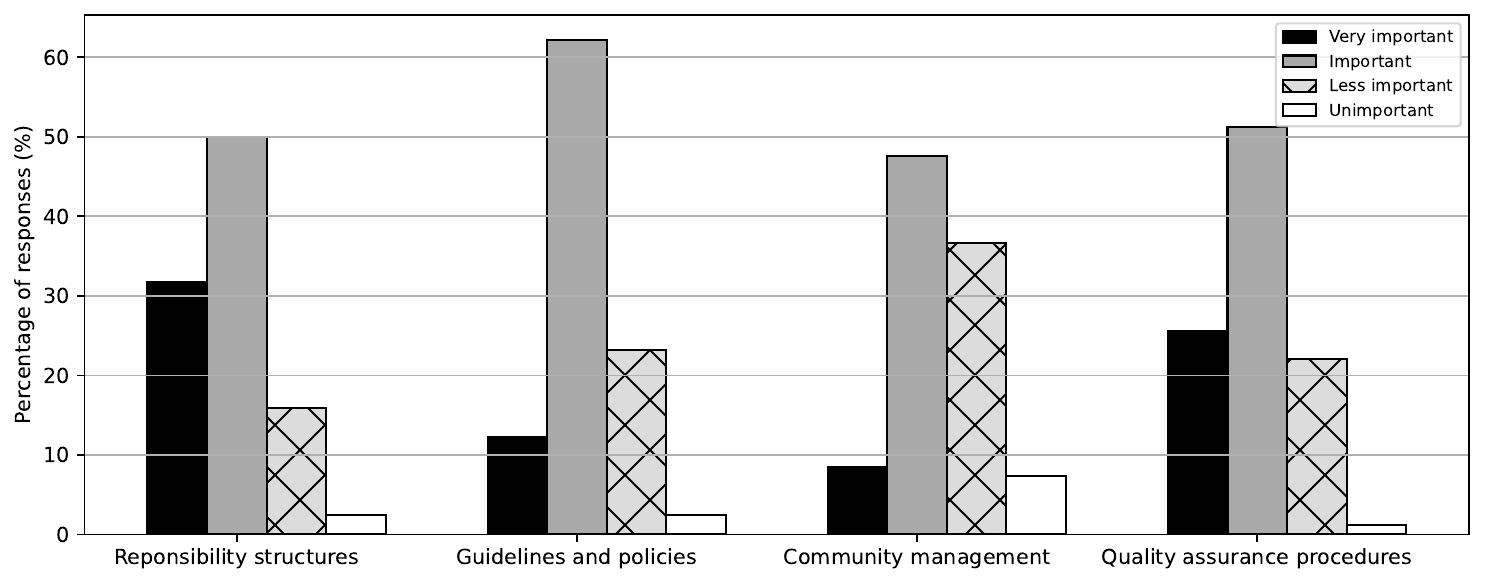}
    \caption{RSE's relevance ratings of organizational aspects for automated research software publication systems}
    \label{fig:preferences_2}
\end{figure}

Responsibility structures for software maintenance were rated as necessary by the majority of participants, with 31.3\% rating it as ``very important'' and 49.4\% as ``important,'' collectively making up 80.7\% of responses (\autoref{fig:preferences_2}). A smaller portion of respondents (15.7\%) rated it as ``less important,'' and only 2.4\% considered it ``unimportant.'' Guidelines and policies for using the system were also highly valued, with 12.0\% rating them as ``very important'' and 61.4\% as ``important,'' totaling 73.4\%. A notable 22.9\% found this aspect ``less important,'' while 2.4\% rated it ``unimportant.'' This result underscores the importance of structured guidelines to support system usage. Community management and building received mixed responses. While 47.0\% rated this aspect as ``important,'' only 8.4\% considered it ``very important''. A significant portion (36.1\%) viewed it as ``less important,'' and 7.2\% considered it ``unimportant.'' This suggests community-related efforts are valued but not as critical as other organizational aspects.

Quality assurance procedures were rated as highly important, with 25.3\% of participants marking them as ``very important'' and 50.6\% as ``important,'' accounting for 75.9\% of responses (\autoref{tab:table3}). Fewer participants rated it as ``less important'' (21.7\%) or ``unimportant'' (1.2\%). These findings emphasize the necessity of robust quality assurance in automated systems. \autoref{fig:preferences_2} shows the distribution of importance ratings across the four organizational aspects, highlighting responsibility structures and quality assurance procedures as the most critical organizational aspects. At the same time, community management received the least prioritization.

Participants subsequently rated whether the roles of RSEs and researchers were clear and defined to RSEs. Results showed that the majority of respondents (77.1\%) characterized the role understanding of RSEs as ``unclear and vague,'' while only 22.9\% rated it as ``clear and defined.''  The participants indicated that they understand the researchers' role slightly better, with 34.9\% rating it as ``clear and defined'' as their role as RSE by 65.1\%  rating the RSE role as ``unclear and vague''.

While structured training sessions are valued by 42.2\% of RSEs as a way to improve collaboration between RSEs and researchers, the majority (57.8\%) did not select this option, indicating that other measures may be more critical. The most highly rated measure was joint project planning and implementation, with 78.3\% of participants selecting this option. Half of the participants (50.6\%) consider better communication and exchange platforms as necessary to improve the mutual understanding of RSEs. This result emphasizes the need for tools and platforms that facilitate interaction and knowledge sharing between RSEs and researchers. 62.7\% of participants considered clear responsibilities and role distribution important. This measure ranks second in preference, reflecting the significance of defining roles and expectations to minimize misunderstandings and foster effective collaboration. Participants also suggested improving the cooperation between RSEs and researchers by establishing more clarity in authorship to ensure that contributions or both are correctly recognized. Participants even suggested that RSEs and researchers should not clearly define deliverables and expectations for collaboration rather than continually redefining desired features or outcomes.
Additionally, participants highlighted the need for more RSE-specific jobs, departments, and institutes and increased resources for researchers who handle significant programming tasks but lack access to RSE support services within their organizations. Finally, they demanded a deeper understanding of the differing goals, such as where researchers produce academic papers and RSEs build functional and maintainable code. Addressing this disconnect could lead to more productive and mutually beneficial collaborations. These suggestions underscore the importance of role flexibility, resource availability, and aligning expectations to strengthen the relationship between RSEs and researchers.

\subsubsection{IF.} 
\begin{figure}[h!]
    \centering
    \includegraphics[width=.8\textwidth]{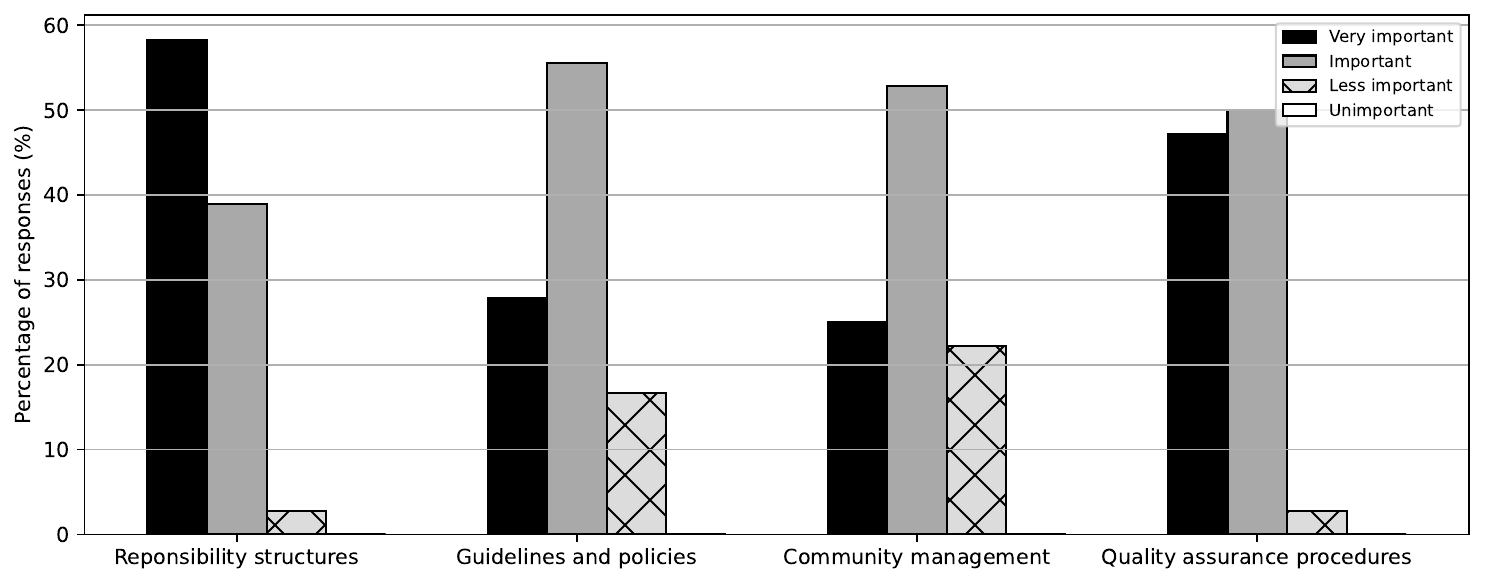}
    \caption{IF's importance ratings of organizational aspects}
    \label{fig:preferences_3}
\end{figure}

As with the RSEs, we asked IFs how they rate the importance of different organizational aspects (see~\autoref{fig:preferences_3}). More than half of the IFs considered responsibility structures for system maintenance as ``very important'' (58.3\%) or ``important'' (38.9\%). The perceived necessity of guidelines and policies for system usage and community management and building showed slightly more variation in the results. 55.6\% rated them as important and 27.8\% as very important. However, 16.7\%  considered them less important. 52.8\% of IFs rated community management and building as important and 25.0\% as very important, while a larger proportion (22.2\%) found it less important. Results thus suggest that community support is valued but not as a relevant organizational aspect for automatic research software publication. Finally, 47.2\% of IFs considered
the application of quality assurance measures in systems for automated software publications ``very important'' while 50\% found them ``important''.

As the the RSEs, IFs describe the role of RSEs with a majority of 74.4\% and of a researcher with 64.4\% as unclear and vague. Only 17.9\% rated the role of the RSE and 28.2\% the role of the researcher as ``clear and defined.'' As a measure to improve collaboration, IF most frequently mentioned clear responsibility and role distribution (28.9\%, N=22) and collaborative project planning (28.9\%, N=22), followed by regular workshops and training (23.7\%, N=18). In contrast, communication platforms were considered less frequent 18.4\% (N=14). Insightful were also the open answers provided. Here, IFs state that RSE and researchers are identical and that the RSE role even does not exist in their organization or university.

\subsection{Capacity-Building Needs}
\subsubsection{RSE.} RSEs most highly endorsed online tutorials and documentation (95.2\%) as the most important learning materials, followed by introductory courses to familiarize users with the platform, which 60.2\% of participants considered as necessary. Nearly half of the participants (48.2\%)  require forums and discussion groups to share their experiences and to find knowledge about the problems at hand.  19.3\% considered advanced workshops on specific functions to be relevant.  In an additional free answer, one participant suggested that the system should be intuitive enough not to require any training, as otherwise, it may deter users from engaging with the platform. This highlights the need for a user-centered design to minimize the need for extensive training.
\subsubsection{IF.} Similar to the answers of RSEs, the results of IFs indicate that online tutorials and documentation were the most frequently selected training measure, with 91.2\% of respondents considering it relevant. Similarly, introduction courses for platform usage were widely regarded as essential by 88.2\% of IF respondents. In contrast, advanced workshops on specific functionalities were chosen less frequently, with 50.0\%, and forums and discussion groups for experience sharing with 32.4\% considering them as relevant. The least frequently selected training option was intensive courses, such as Summer or Winter Schools, chosen by 26.5\% of respondents. 
\subsection{Diversity and Inclusion}
To find out more about the type of diversity (gender, research domain) more relevant to automatic software publication, we also elicited the importance of promoting gender diversity within the Research Software Engineer (RSE) community and evaluated specific measures to enhance the participation of non-male individuals. Participants consider promoting gender diversity in the RSE community as ``very important'' 31.3\% and  ``important''  (36.1\%), collectively making up 67.5\% of responses. Meanwhile, 12.0\% rated it as ``less important,'' and 20.5\% considered it ``unimportant.'' Mentoring programs or networking events for non-male individuals were rated as ``very important'' by 26.5\% and ``important'' by 44.6\%, with a combined 71.1\% supporting such initiatives. Visibility and recognition of the achievements of non-male individuals were rated as ``very important'' by 44.6\% and ``important'' by 32.5\%, making it the most strongly supported measure, with 77.1\% approval. Workshops and training sessions for the community to raise awareness about gender diversity were rated as ``very important'' by 15.7\% and ``important'' by 24.1\%, with 39.8\% approval. Workshops promoting gender diversity received the least support, with only 10.8\% rating them as ``very important'' and 32.5\% as ``important,'' for 43.4\%. A significant proportion of participants (34.9\%) rated this measure as ``less important,'' and 21.7\% rated it as ``unimportant.''

\section{Discussion and Conclusion}

Results show different technical, organizational, and social challenges in designing research infrastructure software for automated software publication, with differing importance ratings across stakeholder groups.
For RSEs, the most essential requirement (83.1\% ``very important'' answers) that a software system for automated software publication such as HERMES must fulfill is compatibility with existing infrastructure. At the same time, this is the only requirement that was rated as ``very important'' by more than 50\% of respondents. In contrast, less than half (48.7\%) of responses from IFs rate this aspect as ``very important''. Vice versa, while documentation -- both for users and administrators -- is very important to IFs (64.1\% and 59\% ``very important'' answers respectively), the requirement for documentation that has been determined ``very important'' by RSEs the most is ``Availability of how-to guides'', with only 36.1\%. Similarly, while IFs assess the importance of usability for system users the highest (84.6\% ``very important'' answers), only 47\% of respondent RSE users rate ``out-of-the-box usability'' as ``very important''. This difference specifically may show that IFs misjudge their users' requirements to some extent.

In other areas of technical requirements, both stakeholder groups provide similar importance ratings, e.g.,
when it comes to compliance of the system with metadata standards, which around half of both IFs and RSEs (53.8\& and 45.8\% respectively) rate as ``very important'', and more than 80\% (IFs: 87.1\%; RSEs: 81.9\%) rate as either ``very important'' or ``important''.

Generally, the discrepancies in weighting technical requirements pose a challenge for the structured validation of requirements and the subsequent user-centric design of the HERMES system. One strategy for handling these discrepancies would be to 
a) deprioritize requirements that rank high in one but not the other stakeholder group, and 
b) prioritize requirements that both stakeholder groups rate as at least ``important''.

In terms of organizational and social aspects, discrepancies in ratings are present across almost all surveyed factors.
Notably, IFs consistently rated all of the following more to be ``very important'' than RSEs: ``Responsibility structures'' (58.3\% vs. 31.3\%), ``Guidelines and policies'' (27.8\% vs. 12\%), ``Community management'' (25\% vs. 8.4\%) and ``Quality assurance procedures'' (47.2\% vs. 25.3\%).
Analogously, only RSEs considered any of the aspects ``unimportant'' at all. This speaks to fundamentally different foci of the stakeholder
groups with respect to organizational and social factors in research infrastructure software systems.

A superordinate underlying challenge that affects the design of HERMES is that only half of respondents practice software publication as defined in~\cite{druskat_towards_2023}.
It is unclear if this is a social, technological, or organizational challenge, and further research is needed to clarify why.
As the practice of FAIR software publication is influenced by cultural aspects within different research communities,
this challenge may be addressed through culture change, whose consideration is outside the scope of this study.

Generally, the internal heterogeneity, especially within the RSE population, poses further challenges.
Based on our results, RSEs demonstrate varied approaches in terms of software publication.
Their research backgrounds span disciplines such as natural sciences, computer science, engineering, social sciences, and the humanities; each coming with distinct knowledge backgrounds and thus different requirements towards research software publication. When designing research infrastructure software, domain-specific practices and cultural differences should be considered.
Similarly, RSE respondents indicated widely differing experiences with research software engineering.
As a result, infrastructure software, related organizational processes, and capacity-building measures should be designed to accommodate these different experience backgrounds, which affects usability, feature sets, and user interfaces.

In summary, the survey results underscore the need for structured, accessible, and automated systems for research software publication that are usable for technical and non-technical user groups, emphasizing domain diversity.
The following two main factors constitute general challenges for designing research infrastructure software:

\begin{enumerate}
    \item The existence of multiple stakeholder groups whose roles are reciprocally engaged in provider-user relationships with respectively differing contexts and requirements on the system.
    \item The internal heterogeneity of each stakeholder group along multiple dimensions, e.g., technical experience, use, and provision of specific infrastructure systems, preferences for and experience with different capacity-building measures, etc.
\end{enumerate}

The present study marks a pivotal improvement in our understanding of the complex interrelations involved in designing research infrastructure software within multi-stakeholder environments. Pinpointing critical aspects and misalignments between RSEs and IFs, our findings mark a first step into the elicitation of user requirements and the user-centered development of systems like HERMES, ensuring their acceptance and widespread use of the interdisciplinary user base.
Adopting participatory design approaches and facilitating robust cross-disciplinary collaborations are imperative to fostering future advancements in automatic research software publication.

The survey used Likert scales to elicit attitudes towards automated software publication systems.
While these scales help express priorities, they are not optimal for a comparative analysis of requirements across stakeholder groups.
To better express differing attitudes towards specific requirements,
both stakeholder groups could have been given the same set of options in terms of technical, social, and organizational requirements,
and could have been asked to rank rather than rate them.
This would have made differences in attitudes clearer.
Similarly, we cannot generalize preference for automated software publication solutions in the RSE population,
as it was elicited conditionally only for respondents who already practice software publication.
This excluded around half of the population from the results.
Similarly, we cannot analyze reasons for not publishing software from RSEs,
as the survey included no questions to that effect.
More generally, the survey did not ask for rationales for answers, which may have led to less comparable results.
Finally, RSEs from the social sciences, the humanities, and medical sciences were underrepresented
in respondents.
This may lead to results that overfit for the more highly represented parts of the population.
Future work should address these limitations.

\begin{credits}
\subsubsection{\discintname}
The authors have no competing interests to declare that are relevant to the content of this article.
\end{credits}

\bibliographystyle{splncs04}
\bibliography{references}
\end{document}